# Discovery of ideal Weyl points with helicoid surface states

Biao Yang[1†], Qinghua Guo[1,4†], Ben Tremain[2†], Rongjuan Liu[3†], Lauren E. Barr[2], Qinghui Yan[3], Wenlong Gao[1], Hongchao Liu[1], Yuanjiang Xiang[4], Jing Chen[5], Chen Fang[3], Alastair Hibbins[2]*, Ling Lu[3]*, Shuang Zhang[1]*

[1] School of Physics and Astronomy, University of Birmingham, Birmingham B15 2TT, United Kingdom.

[2] Electromagnetic and Acoustic Materials Group, Department of Physics and Astronomy, University of Exeter, Stocker Road, Exeter EX4 4QL, United Kingdom.

[3] Institute of Physics, Chinese Academy of Sciences/Beijing National Laboratory for Condensed Matter Physics, Beijing 100190, China

[4] Key Laboratory of Optoelectronic Devices and Systems of Ministry of Education and Guangdong Province, College of Optoelectronic Engineering, Shenzhen University, Shenzhen 518060, China.

[5] School of Physics, Nankai University, Tianjin 300071, China

*Correspondence to: A.P.Hibbins@exeter.ac.uk; linglu@iphy.ac.cn; s.zhang@bham.ac.uk

†These authors contributed equally to this work.

**Abstract**: Weyl points, serving as monopoles in the momentum space and laying the foundation of topological gapless phases, have recently been experimentally demonstrated in various physical systems. However, none of the observed Weyl degeneracies are ideal: they either offset in energy or coexist with trivial dispersions at other momenta. The lack of an ideal Weyl system sets a serious limit to the further development of Weyl physics and potential applications. Here, by constructing a photonic metamaterial, we experimentally observe an ideal Weyl system, whose nodal frequencies are pinned by symmetries to exactly the same value. Benefitting from the ideal Weyl nodes, we are able to map out the complete evolution of the helicoid surface states spinning around the projections of each Weyl nodes. Our discovery provides an ideal photonic platform for Weyl systems and novel topological devices.

Weyl points are relativistic linear crossings of energy bands, serving as the counterparts of magnetic monopoles in the momentum space [1]. With the unit topological charge $\pm 1$, Weyl degeneracies are the basic elements in constructing topological gapless phases, such as double Weyl points [2], Dirac points [3], *spin - 1* Weyl points, etc [4,5]. Up to date, Weyl degeneracies of various forms have been proposed in nearly all wave systems including condensed matters [2,6-11], photonics [12-16] and acoustics [17]. Among them, surface arc as one of the fingerprints of Weyl systems has been widely observed. However, demonstration of some fundamental topological features of Weyl points, such as helicoidal structure of topological surface states [18], is hindered by the complicated configuration of energy bands at the Weyl energy. Moreover, some realistic and innovative device applications critically depend on the cleanliness and simpleness of Weyl systems [1]. Thus, an ideal Weyl system [19-21], so called three dimension graphene [22], is highly required, because in such a system all Weyl nodes are symmetry-related, residing at the same energy with a large momentum separation, and devoid of non-topological bands in a sufficiently large energy interval.

While Weyl degeneracies can be readily found in systems with either breaking time-reversal or inversion symmetry [1], experimental realization of a truly ideal Weyl system turns out to be very challenging and has not yet been reported until now. Here, by exploring electromagnetic response of metallic structures in the crystalline lattice, termed as meta-crystals, we realize an ideal photonic Weyl system protected by $D_{2d}$ point symmetry. The system exhibits four Weyl points at the same energy, which is the minimum number allowed in the presence of time-reversal symmetry. Through near field scanning measurement of the transmission field upon a point source excitation, we observe the intriguing helicoidal structure of topological surface

states - a physical representation of Riemann surface generated by a multi-valued function [18].
Our design offers an ideal platform with the simplest *k*-space configuration for investigating various unconventional physics in Weyl systems.

The structure of the designed ideal photonic Weyl meta-crystal belongs to the simple tetragonal lattice with symmorphic space group P$\bar{4}$m2 (*No. 115*). In each unit cell, there contains a saddle-shaped connective metallic coil (Fig. 1a and 1b), which possesses $D_{2d}$ ($\bar{4}$2m in Hermann-Mauguin notation) point group symmetry. Obviously, the system has no spatial inversion. These metallic elements support localized electromagnetic resonances with current distributions expandable into multipolar modes [23]. In an effective medium model (see supplementary material: Modelled Effective Hamiltonian and Effective Media analysis, Fig. S1), these resonances collectively exhibit bi-anisotropic effect, leading to a directionally dependent chirality response [24]. Here, the unavoidable crossings between the longitudinal mode (LM) with negative dispersion and the transverse modes (TM) with positive dispersion along Γ-M result in the formation of type-I Weyl points as shown in Fig. 1c (for details see Fig. S2) [25]. From irreducible representation analysis of point group, these two modes belong to two different classes with eigenvalues ± 1 of $C_2$ rotation along Γ-M (Fig. 1d), where level repulsion is forbidden (see supplementary material: Eigen electric field and symmetry analysis, Fig. S3). The other three Weyl points can be obtained after symmetry operations of $D_{2d}$. For instance, three two-fold rotation symmetries ($C_2$ and *$2C_2$'*) combined with time-reversal symmetry guarantee that these four Weyl nodes locate on Γ-M at exactly the same frequency, where two mirror symmetries ($\sigma_x$ and $\sigma_y$) can reverse the corresponding topological charges. Figure 1e shows the simulated band structure along high symmetry lines (as defined in Fig. 1d) in the Brillouin zone

(BZ), where a pair of Weyl points resides at the same frequency. Weyl degeneracies appearing in a relatively large energy window (~2.07 GHz around Weyl frequency, cyan shadow region in Fig. 1e) and devoid of other bulk bands greatly facilitate experimental identification.

To confirm the linear band crossings around the Weyl energy, we carried out angle resolved transmission experiment [13]. In order to match the Weyl points to the small in-plane momentum of an incident wave, a sample with special crystal-cutting is fabricated, as shown in Fig. 2a, where the crystal orientation forms an angle of 26.57° with one of the cutting boundaries. Compared with the global axis - *xyz*, a local coordinate - *uvw* is defined. The length (along *u*), width (along *v*) and height (along *w*) of the sample are 300 mm, 100 mm and 300 mm, respectively. Two parameters ($\theta$ and $\varphi$), as illustrated in Fig. 2a and 2b, are scanned to obtain the angle-resolved transmission spectrum. With this specific crystal cutting, Weyl points in BZ are projected along $k_p$, which is indicated in Fig. 2c. Obviously, two of the projected Weyl points are located within the light circle (magenta circle) at the Weyl frequency (13.5 GHz). Thus, even a plane wave illuminated directly from air onto the sample can address these two Weyl points. Comparisons between the simulation and experiment results are shown in Fig. 2d – 2f. In Fig. 2d, with $\varphi = 0°$, linear gapless energy dispersion is obtained and the density of states vanishes at Weyl frequency due to the absence of other bands in energy in an ideal Weyl system. After rotating the sample with 30°/ 60° ($\varphi = 30°/ 60°$) along *v* axis, a complete gap is observed as expected.

Another direct manifestation of the topological aspects of Weyl system is the exotic topological surface states taking the form of arcs connecting between topologically distinct bulk states. Following a closed contour around an end of the arcs, one moves between the valence and conduction bands [18], which is the direct consequences of the chiral characteristic of Weyl nodes as schematically shown in Fig. 3a. It has been known that the gapless surface states of Weyl crystals take the form of helicoid Riemann surfaces [18], where the bulk Weyl points correspond to the poles and zeros adopting the sign of their Chern numbers. Recently, it was shown that topological surface states of double-Weyl systems can be analytically expressed, in the entire BZ, as the double-periodic Weierstrass elliptic function [26]. Since Weierstrass elliptic function has one second-order pole and one second-order zero, it is not the most fundamental expression of the Weyl surfaces states. Here, we show that our ideal-Weyl crystal of four Weyl points have surface states topologically equivalent to the Jacobi elliptic function $cn\,(z, m)$ of two poles and two zeros on the complex plane. $cn\,(z, m)$ is a meromorphic function with periods $4K\,(m)$ and $4K\,(1 - m)$, where K is the complete elliptic integrals of the first kind. For our system, the mapping is given by $\omega\,(k_x, k_y) \sim cn\,((k_x - k_y) / 2 + (k_x + k_y)i / 2, 1 / 2)$, as plotted in Fig. 3a.

In order to explore the helicoidal structure of the surface arcs, transmitted near-field scanning configuration with the excitation source located at the bottom center (Fig. 3b, setup 'a') is applied here, where the detecting probe can raster-scan the top surface to map out both the bulk and surface modes. Another configuration (setup 'b' as shown in Fig. S4b), in which the excitation source is positioned at the edge or corner of the top surface, is also employed to better identify the surface states. These two setups provide complementary information for the observation of helicoid surface states. In all near field measurements, we set the scanning step as 1 mm ($a\,/\,3$), providing a large surface momentum space in the range $[-3\pi\,/\,a,\,3\pi\,/\,a]^2$ after the Fourier transform.

As shown in Fig. 3c, at 13.1 GHz which is below the Weyl frequency, the Fourier transform of the experimentally measured field distribution shows the presence of four symmetrically displaced elliptical bulk states with the same size located along the diagonal directions. We clearly observe two surface arcs running across the BZ boundaries and connecting between the neighboring bulk states with opposite topological charges. In the vicinity of the middle air equi-frequency contour (air circle), there exists a surface ellipsoid. As will be shown later, the surface ellipsoid joins and reroutes the surface arc at higher frequencies. Indeed, the surface ellipsoid and surface arcs together form the same unified helicoid surface in the surface state dispersion.

The helicoid structure of surface arc is investigated by measuring and numerically simulating a series of equi-frequency contours (EFC) between 12.6 GHz and 14.0 GHz, as shown in Fig. 3c and 3e in experiment and Fig. 3d and 3f in simulation, respectively. With increasing frequency, the top surface arc emerged from the Weyl node with positive/negative topological charge rotates anti-clockwise/clockwise. The observed rotation of the helicoid surface state around a Weyl node can therefore be used to detect the chirality of the Weyl node [27]. At lower frequencies, as mentioned above each surface arc connects between the bulk states through the BZ boundary, while the surface ellipsoid expands gradually with frequency. Between 13.5 and 13.6 GHz, the surface arc and surface ellipsoid connect into each other, and then reroute into a new configuration: a direct surface arc connecting between the bulk states within the BZ, and a surface ellipsoid centering at the edge of BZ. The evolution of the surface arc configuration across the measured frequency range matches topologically with that described by the Jacobi elliptic function shown in Fig. 3a. At the frequency of 14.3 GHz, the surface arcs appear to be quite linear (see Fig. S4f), leading to nearly diffraction-less propagation of the surface wave in the real space (see Fig. S4c). Slightly away from the Weyl frequency, the EFC of the bulk state consists of four very small spheres enclosing the Weyl points. It is expected that the interference between them show chessboard like interference pattern in the real space, which is experimentally confirmed (see Fig. S5).

Having confirmed the helicoid nature of the surface state arcs, we now look into the dispersion of the bulk and surface states along four representative momentum cuts, which provide further insight into the bulk-surface correspondence hosted by the ideal Weyl meta-crystal. Two of the band dispersions are taken along two straight momentum cuts passing through the Weyl points in the horizontal and vertical direction, by fixing $k_y = 0.4\pi / a$ (cut-I) and $k_x = 0.4\pi / a$ (cut-III), respectively (Fig. 4c). As shown in Fig. 4, along the horizontal momentum cut, it is observed that a surface state connects directly between the two Weyl points within the BZ (Fig. 4a, cut-I), which is reminiscent of the bearded edge state connection between the Dirac points in graphene. On the other hand, along the vertical momentum cut, the two Weyl points are connected by a surface state crossing the boundary of BZ (Fig. 4a, cut-III), which reminds of the edge state connection in graphene along the zigzag edge [22,28]. For the other two momentum cuts slightly tilted from the horizontal and vertical directions, since they do not pass through the Weyl points, a complete band gap opens up in each band structure. In each case, it is observed that a single surface state connects between the upper and lower bulk bands across the bandgap, indicating the nontrivial topology of the band structures. The numerically simulated dispersion relation along all the four momentum cuts (Fig. 4b) agree very well with the experimental results (Fig. 4a), further revealing the intriguing bulk surface correspondence present in the Weyl system.

In summary, we have designed and observed an ideal Weyl system which had not been realized previously in any experiments. In comparison to the non-ideal ones, the ideal Weyl system opens up direct opportunities for studying the intriguing physics and offers prototype platform for realistic device applications. In photonics, besides the topologically nontrivial surface states supported by the Weyl materials, the diverging Berry curvature close to the Weyl points provides a new degree of freedom in controlling the transport of optical wave packet and may lead to observation of gigantic Hall effect of light [29]. Furthermore, any phenomenon related to the conical dispersion of light cone may be observed around Weyl energy, such as diverging and diminishing scattering cross sections [30]. The vanishing density of states at Weyl frequencies also provides a robust platform for controlling light matter interaction when emitters are embedded inside the photonic Weyl materials. Ideal Weyl materials also open gate to explore some very exceptional electromagnetic phenomena, such as negative refraction and momentum filter [21].

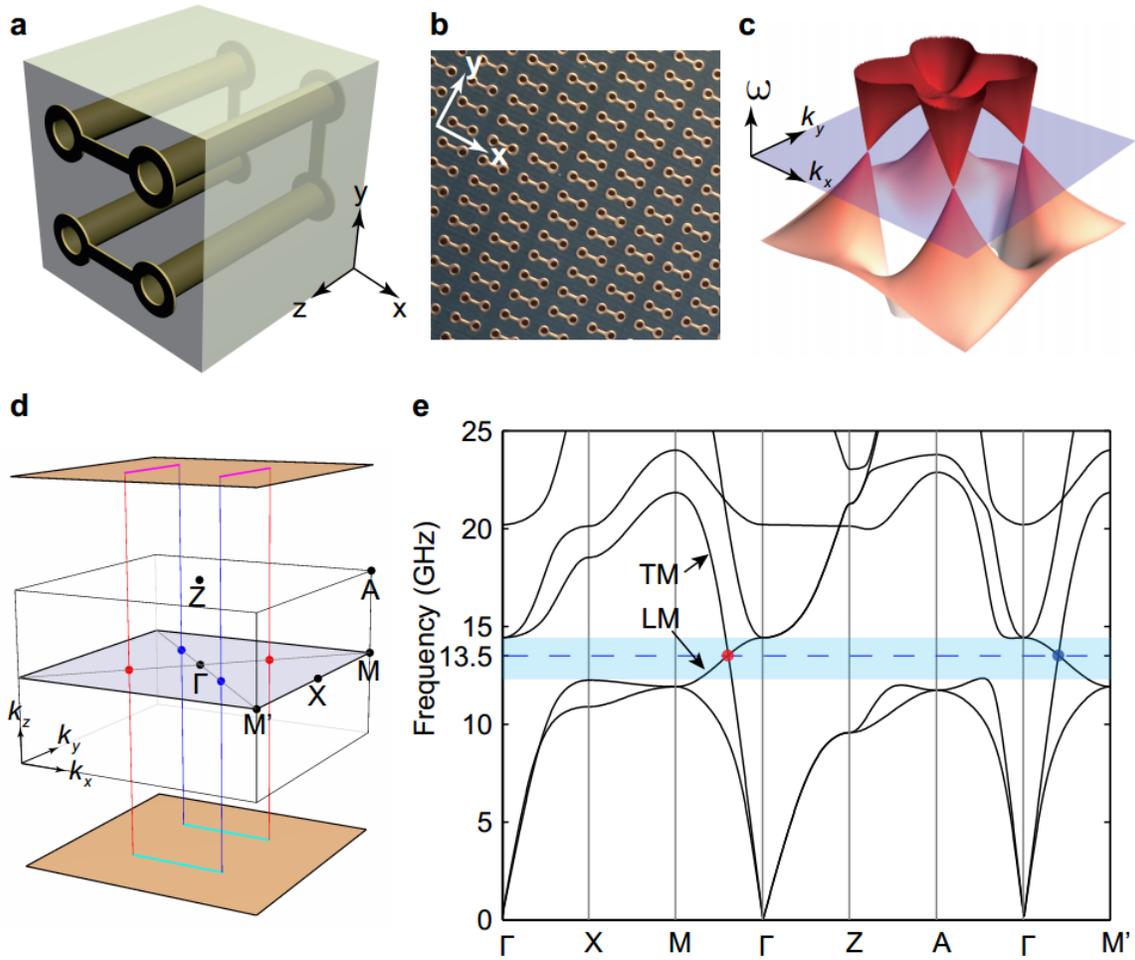

**Fig. 1. Structure and band topology of the ideal photonic Weyl meta-crystal.** (**a**) Unit cell consisting of saddle shaped metallic connective structure, which has non-centrosymmetric $D_{2d}$ point group symmetry. The space group is *No. 115*. (**b**) The top surface of the sample is fabricated with printed circuit board technology. The bulk sample is assembled as layer-by-layer stacking along *z* direction. There is a 1.5 mm-thick blank layer (not shown) between two adjacent 3 mm-thick structure layers to prevent short contacting from metallic coils. Consequently, the saddle coil structure can be viewed as periodically buried in a background material with dielectric constant of 2.2 where the periods along *x*, *y* and *z* directions are $a_x = a_y =$

$a$ = 3 mm and $a_z$ = 4.5 mm, respectively. (**c**) Four type-I Weyl points reside on the same energy as indicated by the blue plane with respect of $k_z$ = 0. (**d**) Bulk and surface Brillouin zone with four Weyl points located on Γ-M. Top and bottom topological surface-state arcs are schematically shown. (**e**) CST simulated band structure along high-symmetry lines. Cyan shadow region highlights the clean window where Weyl points (red/blue point) reside. Longitudinal mode (LM) and transverse mode (TM) are indicated.

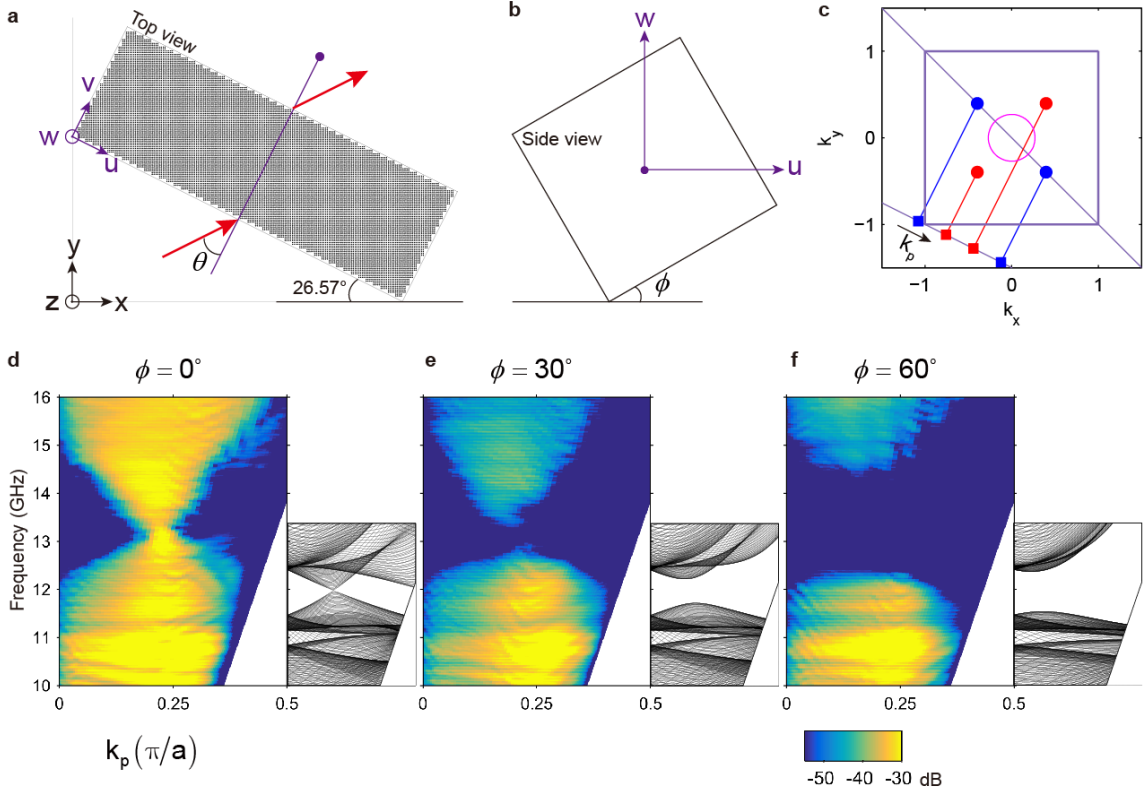

**Fig. 2. Angle resolved transmission measurement of the ideal Weyl system.** (**a**) and (**b**) Schematic view of the sample fabricated with special crystal cutting. Top and side views are indicated. $\theta$ and $\varphi$ are rotation angles defined along the local coordinates *w* and *v*, respectively. (**c**) Projection of Weyl points in momentum space with respect of the global coordinates (*x, y, z*). First Brillouin Zone is indicated by the purple square. Magenta circle indicates the equi-frequency contour of vacuum at 13.5 GHz. (**d**), (**e**) and (**f**) are the band projections with $\varphi = 0°$, 30° and 60°, respectively. Experiment and simulation results are shown in left and right panels, respectively.

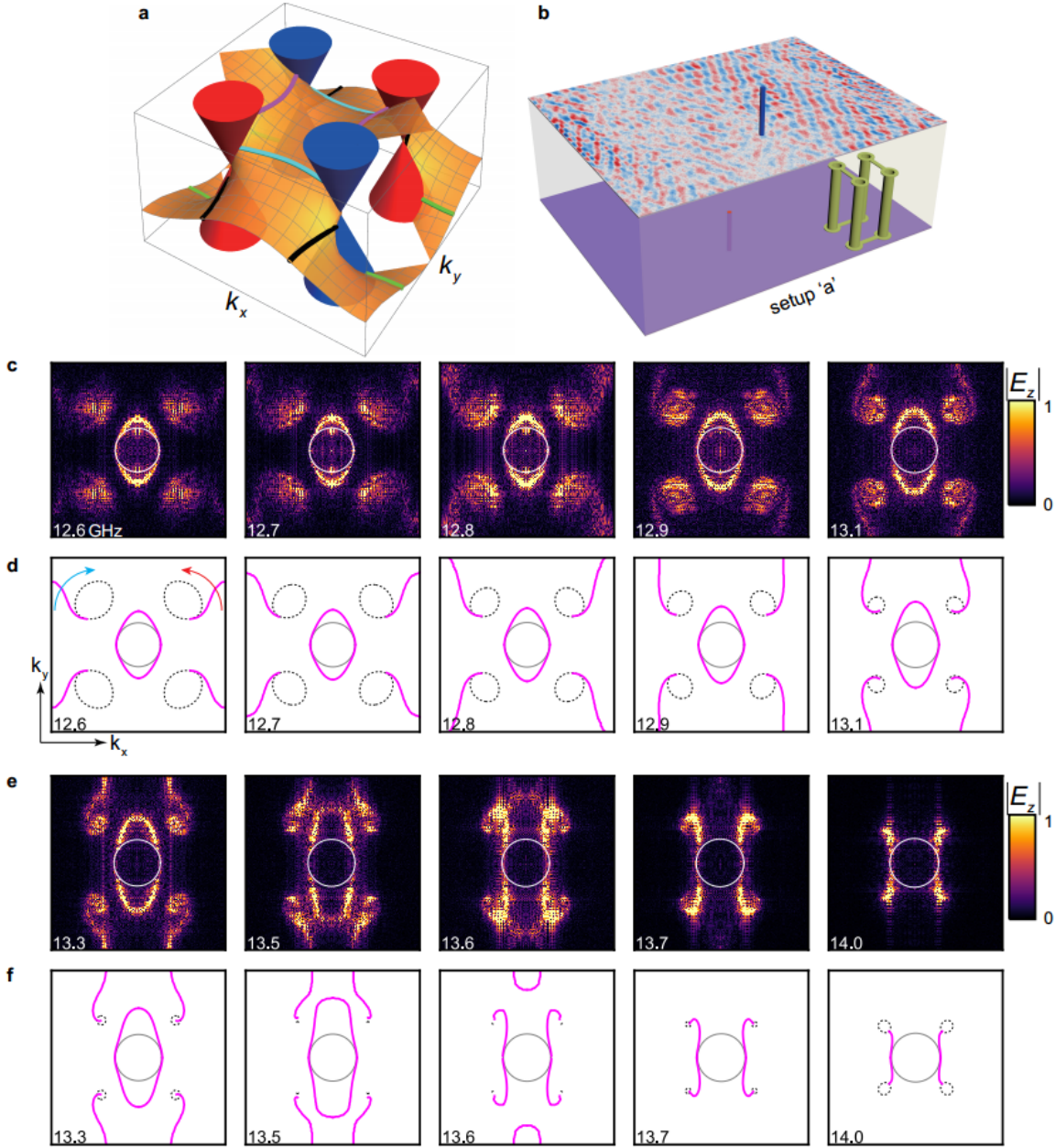

**Fig. 3. Experimental observation of helicoidal structure of topological surface states. (a)** Schematic view of helicoid surface states in an ideal Weyl system with four Weyl points, which is periodic in the Brillouin zone. **(b)** Transmitted near field scanning system (setup 'a'), where the source is positioned on the bottom surface center. **(c)** and **(e)** Equi-frequency contour (EFC) ($|E_z|$) measured under setup 'a' from 12.6 GHz to 14.0 GHz. **(d)** and **(f)** Bulk (black dashed) and surface (magenta solid) states simulated by CST microwave studio, correspondingly. Anti-

clockwise (red) and clockwise (cyan) arrows indicate surface arc rotation directions with increase of frequency corresponding to positive and negative Weyl nodes, respectively. Central solid circle indicates the air EFC. The plotted range for each panel is $[-\pi/a, \pi/a]^2$.

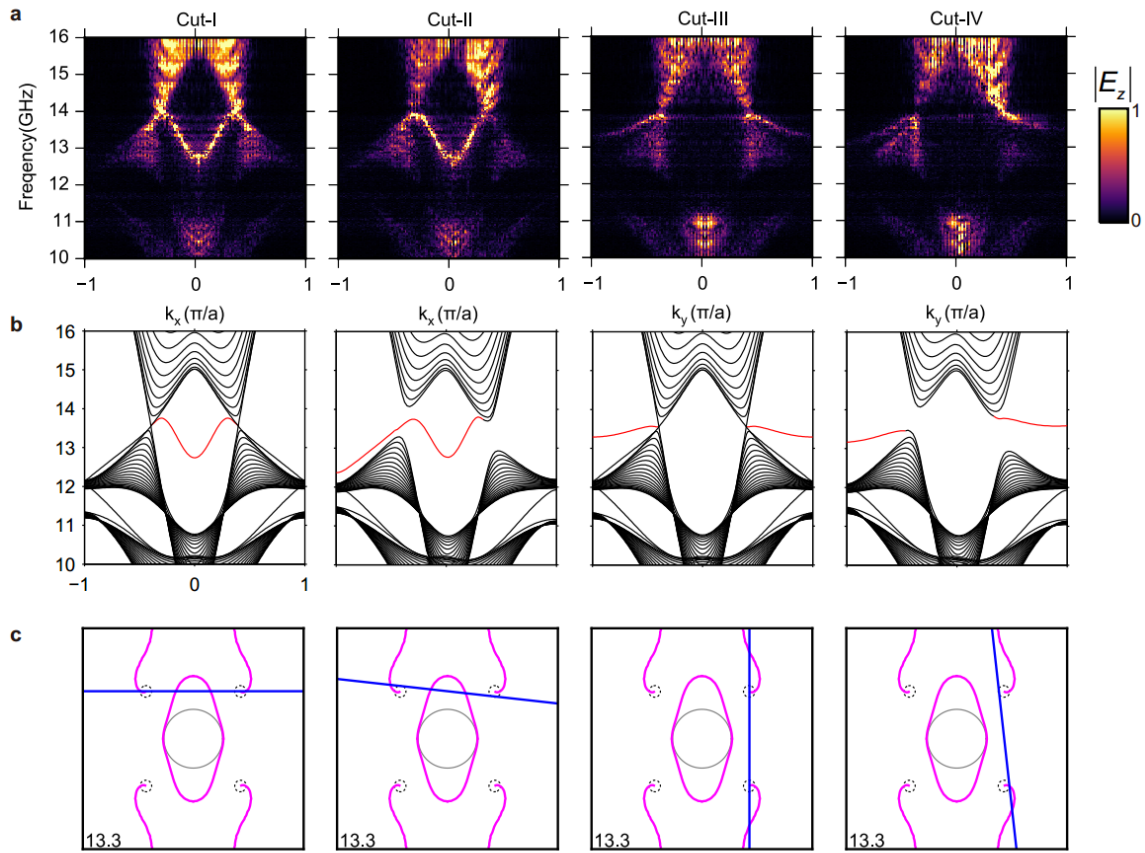

**Fig. 4. The measured and simulated dispersions along four different momentum cuts.** (**a**) The measured band structures along the momentum cuts indicated by (**c**). (**b**) The simulated band structures where the surface states are drawn in red. Surface states in cut-I and cut-III show strong resemblance to the bearded and zigzag edge states of graphene nanoribbon, respectively. (**c**) The numerically calculated equi-frequency contour at 13.3 GHz where the momentum cuts (in blue) for the dispersions shown in (**a**) and (**b**) are indicated.